\documentclass[12pt,twoside]{article}
\usepackage{psfig}
\newcommand{\VolumeHeader}{}

\newcommand{\LectureHeader}{Gravitational Radiation from GRBs}
\begin{document}
\pagestyle{myheadings}
\markboth{\LectureHeader}{\VolumeHeader}
\markright{\VolumeHeader}

%%%%%%%%%%%%%%%%%%%%%%%%%%%%%%%%%%%%%%%%%%%%%%%%%%%%%%%%%%%%%%%%%%%%%%%%%%%
%%%            Title page starts here                                     %
%%%%%%%%%%%%%%%%%%%%%%%%%%%%%%%%%%%%%%%%%%%%%%%%%%%%%%%%%%%%%%%%%%%%%%%%%%%

\begin{titlepage}

%%% YOUR CHANGES BELOW THIS LINE

\title{Gravitational Radiation from Gamma-Ray Bursts\footnote{Lecture given at:
{\it Gravitational Waves: A Challenge to Theoretical Astrophysics Trieste, 5-9 June 2000}}}

\author{Tsvi Piran\thanks{tsvi@nikki.phys.huji.ac.il}
\\[1cm]
{\normalsize
{\it The Racah Institute of Physics, Hebrew University,
  Jerusalem 91904, Israel.}}
}

%%% FOR FURTHER AUTHORS SEE WHAT IT IS WRITTEN IN THE ABSTRACT
%%% DO NOT CHANGE THE FOLLOWING LINES
%{\normalsize {\it Lecture given at the: }}
%\\
%\ActivityName
%\\
%\ActivityDate
%\\[1cm]
%{\small \VolumeSerial}
%\date{}
\maketitle

\baselineskip=14pt
%\newpage
%\thispagestyle{empty}

%%%%%%%%%%%%%%%%%%%%%%%%%%%%%%%%%%%%%%%%%%%%%%%%%%%%%%%%%%%%%%%%%%%%%%%%%%%
%%%            Abstract page starts here                                  %
%%%%%%%%%%%%%%%%%%%%%%%%%%%%%%%%%%%%%%%%%%%%%%%%%%%%%%%%%%%%%%%%%%%%%%%%%%%

\begin{abstract}

Gamma Ray Bursts (GRBs) are the most relativistic objects known so
far, involving, on one hand an ultra-relativistic motion with a
Lorentz factor $\Gamma > 100$ and on the other hand an accreting
newborn black hole.  The two main routes leading to this scenario:
binary neutron star mergers and Collapsar - the collapse of a rotating
star to a black hole, are classical sources for gravitational
radiation. Additionally one expect a specific a   gravitational
radiation pulse associated with the acceleration of the relativistic
ejecta. I consider here the implication of the observed rates of GRBs
to the possibility of detection of a gravitational radiation signal
associated with a GRB. Unfortunately I find that, with currently
planned detectors it is impossible to detect the direct gravitational
radiation associated with the GRB. It is also quite unlikely to detect
gravitational radiation associated with Collapsars. However, the detection of
gravitational radiation from a neutron star merger associated with a
GRB is likely.
\end{abstract}

\vspace{6cm}


%{\it PACS numbers:}

\end{titlepage}

%%%%%%%%%%%%%%%%%%%%%%%%%%%%%%%%%%%%%%%%%%%%%%%%%%%%%%%%%%%%%%%%%%%%%%%%%%%
%%%       Automatic TOC and your Text starts here                         %
%%%%%%%%%%%%%%%%%%%%%%%%%%%%%%%%%%%%%%%%%%%%%%%%%%%%%%%%%%%%%%%%%%%%%%%%%%%

%\newpage
%\thispagestyle{empty}
%\tableofcontents

\newpage

\section{Introduction}

Gamma Ray Bursts (GRBs) are the most relativistic objects known so
far.  According to the fireball model \cite{Piran99,Piran2000} the observed
gamma-rays are produced within relativistic shocks arising when an
ultra-relativistic ejecta (with a Lorentz factor $\Gamma > 100$) is
slowed down.  The GRB arises due to internal shocks within the
ultra-relativistic flow.  The afterglow is produced when the ejecta is
slowed down by the surrounding medium. The ``inner engine" that
accelerates the relativistic flow and powers the GRB involves in most
models a newly formed black hole surrounded by a thick accretion disk
around it. The ultra-relativistic ejecta is, most likely, in the form
of two jets along the rotation axis of this system.

It is only natural to expect that with these processes GRBs will be
accompanied by ``strong" gravitational radiation signals. We identify
two different stages that can produce gravitational radiation. First,
gravitational radiation can arise when the black hole
forms. Currently there are two competing model for the inner engine: A
binary neutron star merger, in which case we should expect a
``classical" chirping signal of a merger \cite{LIGO-merger}, or a Collapsar,
which involves an asymmetric collapse of a massive star, whose
gravitational radiation signal resembles the one from a rapidly
rotating core collapse \cite{Stark_Piran85}.  In both models we find a
black hole surrounded by a massive accretion disk. The accretion of
this disk onto the black hole powers the GRB.  Some of the accreting
material is accelerated, most likely along the symmetry axis, to an
ultra-relativistic velocities. This ejecta produces the GRB.  In
addition to the gravitational radiation produced during the formation
of the ``central engine" we also expect gravitational radiation from
the acceleration stage in which the relativistic ejecta is accelerated
to an ultra relativistic velocity. This emission is unique to GRBs. It
should appear in association with GRBs regardless of the nature of the
model leading to the GRB.

I examine here the possibility that this radiation signal will be
detected in coincidence with a GRB by future gravitational radiation
detectors.  I discuss in section 2 some general features of GRBs which
serve as a background for the following discussion. In sections 3 and
4 I discuss gravitational radiation from GRBs associated with mergers
(section 3) and Collapsars (section 4). In section 5 I review the
basic features of the fireball model and I calculate the gravitational
radiation emitted during the acceleration phase of relativistic
ejecta. The results are summarized in section 6.

\section{Gamma-Ray Bursts}

Three related features of GRBs are most relevant to our discussion: rate,
beaming and energy. I refere the reader to a recent review \cite{Piran99,Piran2000}
for additional details on GRBs.

\subsection{Isotropic Rates}

The rate of GRBs is calculated by fitting the observed BATSE $Log N /Log S$ (actually
the $C/C_{min}$)  distribution \cite{Piran92,Cohen_Piran95} to theoretical models.  In
this work I use the recent results of Schmidt
\cite{Schmidt99,Schmidt00} who confirmed previous indications that
GRBs follow the star formation rate. Schmidt's most recent
\cite{Schmidt00} estimate for the current local rate $R_{iso}= 2
\times 10^{-9} {\rm Mpc}^{-3} {\rm yr}^{-1}$, where anticipating the
discussion of beaming the subscript {\it iso} indicates that this
quantity is calculates assuming isotropic emission.  The current local
rate, which is of interest to us here is lower than the rate in the
past which was higher when the star formation rate was higher.  This
rate corresponds to one burst per $5 \cdot 10^6$y per galaxy for a galaxy
density of $0.01 {\rm Mpc}^{-3}$.  Note that this rate is larger
by a factor of 10 than Schmidt's previous estimate reported a year
ago \cite{Schmidt99}. This difference could serve as an indication to
the possible error in these numbers. The luminosity of GRBs is highly
variable spanning two orders of magnitude. The average energy released
in a burst (assuming isotropic emission), $\bar E_{iso}$ is
$10^{52}$ergs. The bursts with measured red shifts range from $
z=0.433$, for GRB990712 to $z =4.2$ (excluding $z=0.05$ of
SN1998bw which may or may not be associated with GRB980425), the most
distant one observed so far is at $z=4.5$.  Schmit's analysis suggests
that if the bursts are isotropic than once per year we see a burst as
near as 1Gpc ($z= 0.22 h_{66}$).

\subsection{Short and Long Bursts}

The duration distribution of the bursts divides GRBs to two
populations: long and short bursts according to $T> 2$secs or
$T<2$secs.  About one quarter of the observed bursts are short.  The
$Log N/Log S$ distribution shows clearly that the observed short
bursts are not detected from as far as the long ones.  This means that
on average short bursts are weaker and hence the observed short bursts
are nearer to us that the long ones \cite{Mao_narayan_piran,Katz}.
The best estimate so far suggests that all short bursts are at
$z<0.5$.  So far afterglow was not detected from any short burst and
we don't have a redshift measurement to any short burst hence there is
no independent confirmation of this expectation.  Given about 250
short bursts per year and assuming that they all come from within
$z<0.5$ we find that the current rate of observed short bursts is
$R_{iso-short}= 2 \times 10^{-8} {\rm Mpc}^{-3} {\rm yr}^{-1}$
about ten times larger than the rate of long GRBs.

\subsection{Jets and Beaming}

There are various indications arising from GRB afterglows that GRBs
are beamed with beaming angles ranging from $2^o$ to $20^o$
\cite{beaming1,beaming2,beaming3,beaming4}.
Beaming would have an immediate implications on both
the rate and the energy of these events. With a beaming angles of
$\theta= 0.1 \theta_{0.1} $ the overall current rate of GRBs is
$R_{\theta} = 8 \times 10^{-7} \theta_{0.1}^{-2} {\rm Mpc}^{-3} {\rm
yr}^{-1}$ or a burst per $10^4\theta_{0.1}^{-2} $y per galaxy. With
one burst per year at a distance of $135 \theta_{0.1}^{2/3}$Mpc and a
redshift of 0.03. Note, however, that this "nearby" burst will, most
likely, not be beamed in our direction and hence its gamma-ray
emission won't be observed. It could possibly be detected latter as an
orphan afterglow in the optical or the radio bands if its position
would be known with a good enough accuracy.  The overall energy
$E_\theta$ is reduced of course by exactly the same factor $\bar
E_\theta (\theta^2 /4) \bar E_{iso} = 2.5 \times 10^{49}
\theta_{0.1}^2$ ergs.

It is not known if short bursts are beamed. Assuming that they are beamed
with beaming angles similar to those seen in long bursts we
find that the rate of these bursts is one per $10^3\theta_{0.1}^{-2} $y
per galaxy with one burst per year at a distance of $80
\theta_{0.1}^{2/3}$Mpc.

\section{Mergers}

I consider here both
binary neutron star mergers and black hole-neutron star mergers
under the single category of mergers. These sources
are the ``canonical" sources of
gravitational radiation emission.  Both LIGO and VIRGO  aim in
detecting these sources. Specifically the goal of these detectors is
to detect the characteristic``chirping" signals arising from the
in-spiraling phase of these events. The possibility of detection of
such signals has been extensively discussed (see
e.g. \cite{LIGO-merger}) and we won't repeat this here. Such events
could be detected up to a distance of $\sim 20$Mpc with LIGO I
and up to $\sim 300-600$Mpc with LIGO II.

The detection of the chirping merger signal is based on fitting the
gravitational radiation signal to pre-calculated templets.  Kochaneck
and Piran \cite{Kochaneck_Piran93} suggested that the detection of a
merger gravitational radiation signal would requite a lower S/N ratio
if this signal coincides with a GRB. This could increase somewhat the
effective sensitivity of LIGO and VIRGO to such events.

It is expected that mergers (either binary neutron star or a black
hole-neutron star mergers) produce the short GRBs (see
\cite{Narayan01}).  Considering the isotropic rate of short GRBs
estimated earlier we find that there should be one short burst per
year within $\sim 450$Mpc.  This is just at the sensitivity level of
LIGO II.  As already mentioned it is not clear if short GRBs are
beamed.  If they are beamed, with the same beaming factor as long GRBs
we should expect several hundred mergers events per a single observed
burst.  This would put one merger event per year at $\sim
80\theta_{0.1}^2$Mpc.

The corresponding distances to long GRBs are much longer.  The nearest
(long) GRB detected within a year would be a t 1Gpc. This is far
beyond the sensitivity of even LIGO II.  If GRBs are beamed than the
nearest (long) event would be much nearer, at $135 \theta_{0.1}^2$Mpc,
well within the sensitivity of LIGO II.
However, this burst would be directed away from us. Still a GRB
that is beamed away from us is expected to produce an ``orphan"
afterglow and the gravitational radiation signal could trigger a
search for this afterglow.

\section{Collapsars}

The Collapsar model \cite{woosley,woosley_macfayden} is based on the
collapse of the core of a massive star to a black hole surrounded by a
thick massive accretion disk. The accretion of this disk onto the
black hole, is accompanied by the acceleration of ultra relativistic
jets along the rotation axis and powers the GRB. The jets first have
to punch a hole in the stellar envelope. The GRB forms only after the
jets have emerged from the envelope. Due to the relatively long time
that it takes for the jets to punch a hole in the envelope it is
expected that Collapsars can produce only the long bursts.

As far as gravitational radiation is concerned this system is very
similar to a regular supernova.  Rotating gravitational collapse has
been analyzed by Stark and Piran \cite{Stark_Piran85}. They find that
the gravitational radiation emission emitted in a rotating collapse to
a black hole is dominated by the black hole's lowest normal modes,
with a typical frequency of $\sim 20c^3/GM$. The total energy emitted is:
\begin{equation}
{\Delta E_{GW}} = \epsilon M c^2 =
\rm{min}(1.4 \cdot 10^{-3} a^4, \epsilon_{max})   M c^2 \ ,
\end{equation}
where $a$ is the dimensionless specific angular momentum and
$\epsilon_{max}$ is a maximal efficiency which is of the order ${\rm a~~
few} \times 10^{-4}$.  The expected amplitude of the gravitational
radiation signal, $h$, would be of the order of $\sqrt{\epsilon}
GM/c^2 d$ where $d$ is the distance to the source.  Even LIGO II won't
be sensitive enough to detect such a signal from a distance of 1Gpc or
even from 100 Mpc. Furthermore, this signal would be rather similar to
a supernova gravitational radiation signal.  As regular supernovae
are much more
frequent it is likely that a supernova gravitational radiation signal
would be discovered long before a Collapsar gravitational radiation
signal.

\section{Gravitational Radiation from the GRB}

I turn now to examine the gravitational radiation that would arise
from the GRB process itself.  According to the fireball model the
``inner engine" accelerates a mass  $m = E/\Gamma c^2$ to a Lorentz
factor $\Gamma$. Typical values for
$\Gamma$ are 100 or more (see e.g. \cite{Piran99,Piran2000}).
Then a fraction  of this energy is
converted via internal shocks to the gamma-rays.  The rest of the
energy is dissipated latter while it is slowing done by the
surrounding matter, producing the afterglow.

The most efficient generation of gravitational radiation could take
place here during the acceleration phase, in which the mass is
accelerated to a Lorentz factor $\Gamma$. To estimate this emission we
follow Weinberg's \cite{Weinberg73} analysis of gravitational
radiation emitted from a relativistic collision between two particles.
This estimate is different from the ``Christodoulu effect" of
a permanent signature from a source moving at the speed of light
\cite{Christodoulou91}.
We consider the following simple toy model.  Consider two particles at
rest with a mass $m_0$ that are accelerated instantly at $t=0$ to a
Lorentz factor $\Gamma$ and energy $E$.  Conservation of energy
requires that some (actually most) of the rest mass was converted to
kinetic energy during the acceleration and the rest mass of the
accelerated particle is $m = E/\Gamma = m_0/\Gamma$. Using the formalism
developed by Weinberg \cite{Weinberg73} to estimate the gravitational
radiation generated in particle collisions, we calculate the
gravitational radiation emitted by this system. Prior to the
acceleration the two particles has momenta $m_0 (1, 0,0,0)$. After the
acceleration the particles' momenta are $m \Gamma (1, \pm \vec \beta)$.
The energy emitted per unit frequency per unit solid angle in the
direction at an angle $\alpha$ relative to $\vec \beta$ is:
\begin{equation}
{d E \over d \Omega d \omega} = {G m^2 \beta^2 \over c \pi^2}
\big[   {\Gamma^2 (\beta^2 - \cos^2\alpha)
\over  (1 - \beta^2 \cos^2\alpha)^2} + { \cos^2\alpha \over
\Gamma^2 (1 - \beta^2 \cos^2\alpha)^2} \big]\ .
\label{flux}
\end{equation}
The result is independent of the frequency, implying that the integral
over all frequency will diverge. This non-physical divergence arises
from the non-physical assumption that the acceleration is
instantaneous. In reality this acceleration takes place over a time
$\delta t$, which is of order 0.01sec. This would produce a cutoff
$\omega_{max} \sim 2 \pi / \delta t$ above which Eq. \ref{flux} is not
valid.  The angular distribution found in Eq. \ref{flux} is
disappointing.  The EM emission from the ultra-relativistic source is
beamed forwards into a small angle $1/\Gamma$, enhancing the emission
in the forwards direction by a large factor ($\Gamma^2$). We find here
that the gravitational radiation from this relativistic ejecta is
spread rather uniformly in almost all $4\pi$ directions. Instead of
beaming we have ``anti-beaming" with no radiation at all emitted
within the forward angle $\gamma^{-1}$ along the direction of the
relativistic motion.

Integration of the energy flux over different directions yields:
\begin{equation}
{d E \over d \omega} = {G m^2 \over c \pi^2}
[ 2 \Gamma^2 + 1+  {( 1 - 4 \Gamma^2) \over \Gamma^2 \beta} \rm{tg}^{-1}(\beta)]
\ .\label{energy_flux}
\end{equation}
As expected the total energy emitted is proportional to $E^2=m^2
\Gamma^2$.  Further integration over frequencies up to the cutoff $2
\pi / \delta t$ yields:
\begin{equation}
E  \approx { 2 G m^2 \Gamma^2 \over c \pi \delta t } \ .
\end{equation}

In reality the situation is more complicated than the one presented
here. First, the angular width of the emitted blobs is larger than
$\gamma^{-1}$. The superposition of emission from different directions
washes out the no emission effect in the forward
direction. Additionally according to the internal shocks model the
acceleration of different blobs goes on independently. Emission from
different blobs should be combined to get the actual emission.
Both
effects {\it reduce} the effective emission of gravitational radiation
and makes the above estimate an upper limit to the emission that is
actually emitted.

The gravitational signal is spread in all directions (apart from a
narrow beam along the direction of the relativistic motion the
GRB). It ranges in frequency from $0$ to $f_{max} \approx 100$Hz.  The
amplitude of the gravitational radiation signal at the maximal
frequency, $f_{max} \approx 100$Hz, would be: $ h \approx (GM\Gamma^2
/c^2 d) $.  For typical values of $E=m\Gamma = 10^{51}$ ergs, $\delta
t = 0.01$sec and an optimistic distance of $100$Mpc, $ h \approx 2.5 \cdot
10^{-25}$.  Far below the sensitivity of the planned gravitational
radiation detectors. Even if we consider a burst which is ten times
nearer this "direct" gravitational radiation signal would still be
undetectable.

\section{Conclusions}

GRBs produce gravitational radiation in two phases.  The first is
during the formation of the compact object (most likely a black hole)
believed to be associated with the "inner engine of the GRB. The
second is from the acceleration phase of the ultra relativistic
ejecta.  This second emission is ``directly" associated with the GRB
phenomenon and would occur in any GRB model which is based on the
rather well established fireball concept.  Our conclusions are
somewhat disappointing. In spite of the ultra relativistic nature of
the fireballs arising in GRBs they are not potential sources of
detectable gravitational radiation signals.  Gravitational radiation
may be, however, detected from the stellar processes associated with
the energy generation for GRBs. Here the situation is also
disappointing.  If GRBs are associated with mergers we would expect
that once a year there would be a sufficiently nearby merger that
could be detected.  These chances are larger if mergers are associated
with short GRBs which are weaker and have larger event rate than long
ones. However, because of beaming it is most likely that we won't
observe the associated GRB. Gravitational radiation produced within
the Collapsar model are even more disappointing. This emission is very
similar to the emission expected from regular supernovae.  Typical
distances to these GRBs are hundred Mpcs. The gravitational radiation
from such sources is practically undetectable. With the much larger
rate of regular supernovae it is by far more likely to detect a
gravitational radiation signal from a regular supernova than from a
Collapsar.

I thank the relativistic astrophysics group at Caltech for hospitality
while this article was written. This research was supported by a
US-Israel BSF grant.

%\newpage

\end{document}